\documentclass[prd,aps,twocolumn,superscriptaddress,nofootinbib,10pt,
               floatfix]{revtex4-2}

\usepackage{amsmath,amssymb}
\usepackage{graphicx}
\usepackage{float}
\usepackage{xcolor}
\usepackage{xspace}
\usepackage[colorlinks,citecolor=blue,bookmarks]{hyperref}
\graphicspath{{./}}

\newcommand{\Msun}{\,M_\odot}
\newcommand{\fmi}{\,\text{fm}^{-3}}
\newcommand{\MeVfm}{\,\text{MeV\,fm}^{-3}}
\newcommand{\Rof}[1]{R(#1\,M_\odot)}
\newcommand{\R}[1]{R_{#1}}

\newcommand{\MTOV}{M_{\rm TOV}}
\newcommand{\cs}{c_{s}^{2}}
\newcommand{\chiEFT}{\ensuremath{\chi}\textrm{EFT}\xspace}
\newcommand{\pQCD}{\textrm{pQCD}\xspace}
\newcommand{\ecOf}[1]{\varepsilon_{c}(#1\,M_\odot)}
\usepackage{xcolor}

\begin{document}

\title{Causality alone bounds the maximum radius difference between different-mass neutron stars}
\date{\today}

\author{Aleksi Kurkela}
\affiliation{Faculty of Science and Technology, University of Stavanger, 4036 Stavanger, Norway}
\email{aleksi.kurkela@uis.no}

\author{Tuhin Malik}
\affiliation{CFisUC, Department of Physics, University of Coimbra,
  3004-516 Coimbra, Portugal}
\email{tm@uc.pt}

\begin{abstract}
We investigate how the assumption of a common causal equation of state (EoS) correlates the radii of neutron stars at different masses and thereby reduces the uncertainties inferred from independent observations. We show that causality, anchored only to the chiral effective field theory (\chiEFT) EoS near saturation density, places a closed-form upper bound on the radius difference, $\Rof{2.0}\le 1.16\,\Rof{1.4}-1.1\,\mathrm{km}$. The bound is saturated exactly by a one-parameter family of EoSs that we construct analytically. Imposing this prior-independent causal ceiling on the independent NICER posteriors of PSR~J0437$-$4715, PSR~J0614$-$3329, and PSR~J0740+6620 retains only $7.5\%$ of their joint product distribution and removes the large-radius tail of the PSR~J0740+6620 posterior. Unlike full EoS-informed inferences, our construction cleanly isolates the consequences of the generic physical assumptions of a common causal EoS from those associated with a particular choice of EoS prior, providing a transparent benchmark for interpreting neutron-star observations.
\end{abstract}

\maketitle
\section{Introduction}
The matter in a neutron star is assumed to be in the ground state of nuclear matter in weak equilibrium at high density. Consequently, all neutron stars are described by a universal equation of state (EoS), $p(\varepsilon)$, and, through hydrostatic equilibrium, a universal relation between masses, radii, and tidal deformabilities. Current observations probe this relation up to densities of several $n_s$: the two-solar-mass pulsars~\cite{NANOGrav:2019jur,Fonseca:2021wxt,Antoniadis:2013pzd}, the tidal-deformability bound from GW170817~\cite{LIGOScientific:2018cki}, and the NICER radius measurements of PSR~J0030+0451~\cite{Miller:2019cac,Riley:2019yda, Vinciguerra:2023qxq}, PSR~J0740+6620~\cite{Miller:2021qha,Riley:2021pdl,Salmi:2024aum, Dittmann:2024mbo}, PSR~J0437$-$4715~\cite{Choudhury:2024xbk}, and the low-mass pulsar PSR~J0614$-$3329~\cite{Mauviard:2025dmd}. 

Although these measurements are statistically independent, they probe the same underlying EoS. As a result, the radii of neutron stars at different masses are linked. A simple observable capturing this connection is the radius difference
$\Delta R=\Rof{2.0}-\Rof{1.4}$, which compares stars whose structures are governed by different density regimes of the common EoS. The quantity $\Delta R$ has recently been proposed as a probe of high-density stiffening, softening, and possible phase transitions in dense matter~\cite{Drischler:2020fvz,Lin:2023cbo,Tang:2025xib,Ferreira:2024hxc}. Yet all existing inferences involving $\Delta R$ are tied, either explicitly or implicitly, to a particular EoS model or prior. It is therefore natural to ask to what extent $\Delta R$ is constrained by causality alone, independent of any particular EoS parametrization or prior.

The answer is non-trivial because $\Rof{1.4}$ and $\Rof{2.0}$ are sensitive to overlapping, but not identical, density regimes of the EoS. Fixing $\Rof{1.4}$ constrains the low-density EoS sampled by both stars, but a $2 M_\odot$-star also probes substantially higher densities, where neither chiral effective field theory (\chiEFT) nor perturbative QCD (\pQCD) applies directly and phenomena such as strong phase transitions or sound-speed peaks may occur~\cite{Annala:2019puf,Annala:2023cwx,Drischler:2020fvz}. To maximize $\Delta R$, one must identify not only the stiffest causal continuation at high density, but also the low-density EoS from which that continuation emerges. While the maximal $\Rof{2.0}$ is obtained by a $c_s^2=1$ extension, the key challenge is to determine which low-density EoS reproduces a given $\Rof{1.4}$ and at the same time permits the largest possible $\Rof{2.0}$.

In this work, we determine the causal ceiling $R_{2.0}^{\max}(R_{1.4})$ and identify the EoSs that saturate it. We do so using two complementary approaches: a maximally agnostic numerical construction based on the constrained fractal-bridge prior \cite{Gorda:2025aiu} and an analytic characterization of the saturating EoSs and show that they lead to the same ceiling. For comparison, we also study the widely used sound-speed parametrization \cite{Annala:2019puf} and demonstrate that, owing to its limited flexibility, it does not fully saturate the causal ceiling.

The causal ceiling has a direct observational application. Recent analyses have begun to incorporate equation-of-state information directly into the analysis of X-ray pulse-profile observations, rather than treating radius measurements and EoS inference as separate steps~\cite{Hoogkamer:2025eaq}. By enforcing consistency with a shared EoS across multiple neutron stars, these approaches can substantially reduce the posterior uncertainties of the inferred masses and radii. The resulting constraints, however, necessarily inherit the assumptions of the adopted EoS model. Here we show that the model-independent bound on $\Delta R$ derived in this work can achieve a comparable reduction in uncertainty without performing a full EoS inference and without introducing any additional EoS parametrization beyond causality and the assumption of a universal EoS. Applied to the independent NICER posteriors of PSR~J0437$-$4715 and PSR~J0614$-$3329 (low-mass anchors) together with PSR~J0740+6620 (high-mass anchor), the bound removes the causality-incompatible large-radius tail of PSR~J0740 and retains only $7.5\%$ of the joint posterior volume. Furthermore, we show that the resulting allowed region, when combined with the corresponding conformal-sound-speed construction, implies that the squared sound speed in the densest stellar cores must exceed the conformal value $c_s^2=1/3$.

\section{Methodology}
In the following, we describe the components of our EoS construction and the methodology used to determine the causal ceiling.
\paragraph*{Outer crust.}
We adopt the Baym--Pethick--Sutherland (BPS) EoS~\cite{Baym:1971pw} below $n \lesssim 0.5\,n_s$ (with $n_s = 0.16\fmi$).
\paragraph*{Low-density anchor.}
Between $0.5\,n_s$ and $1.1\,n_s$ we match to polytropes that span the \chiEFT\ band of Hebeler~\textit{et~al.}~\cite{Hebeler:2013nza}, with polytropic index $\Gamma$ sampled uniformly in $[1.77,\,3.23]$ to bracket the soft and stiff \chiEFT\ limits. The EoS is thus specified up to a chemical potential $\mu_L \in [0.966,\,0.978]\,$GeV at the \chiEFT\ endpoint $n_L = 0.176\fmi$, with pressure $p_L \in [2.163,\,3.542]\MeVfm$.

\paragraph*{High-density anchor.}
At $\mu_H = 2.6\,$GeV we impose the \pQCD\ result of Fraga~\textit{et~al.}~\cite{Fraga:2013qra}, which translates to $n_H \in [6.14,\,6.87]\fmi$ and $p_H \in [2334,\,4284]\MeVfm$.

\paragraph*{Agnostic ensembles.}
Between the two anchors we sample the allowed EoS space using two complementary constructions. \emph{(i)~Fractal-bridge prior}---we employ the self-similar refinement of Gorda~\textit{et~al.}~\cite{Gorda:2025aiu}, based on the thermodynamic bounds of Ref.~\cite{Komoltsev:2021jzg}. The construction hierarchically samples an increasing number of EoS points, with each refinement layer constrained to remain thermodynamically consistent with all previously sampled points; here we use ten refinement layers, corresponding to $2^{10}+1=1025$ EoS points and yielding a prior containing $N_{\rm fr}\simeq8.6\times10^5$ EoSs. The fractal construction samples the entire causally and mechanically allowed $(\mu,n,p)$ volume and therefore provides the broadest model-agnostic prior consistent with the low- and high-density anchors. Unlike Ref.~\cite{Gorda:2025aiu}, we do not apply diffusive filtering. Consequently, we refer to the construction as the \emph{fractal bridge} rather than the \emph{Gaussian-process bridge} as in \cite{Gorda:2025aiu}. \emph{(ii)~Sound-speed parametrization}---we use the widely used $\cs(\varepsilon)$ construction of Annala~\textit{et~al.}~\cite{Annala:2019puf}, with seven piecewise-linear segments, yielding $N_{c_s^2} \simeq 10^6$ EoSs.

\paragraph*{TOV solution and astrophysical filter.}
{For each EoS we integrate the Tolman--Oppenheimer--Volkoff (TOV) and tidal equations on a dense grid of central energy densities, obtaining the full $M(\varepsilon_c)$ and $R(\varepsilon_c)$ sequence. We identify the maximum-mass configuration $\MTOV$ and discard any EoS that fails $\MTOV > 1.97\Msun$, consistent with the heaviest precisely measured pulsar~\cite{NANOGrav:2019jur,Fonseca:2021wxt}. On the surviving EoSs, $\R{1.4}$, $\R{2.0}$ and $\Lambda_{1.4}$ are read off by interpolation on the stable branch (the branch on which $M$ increases with $\varepsilon_c$ up to $\MTOV$). No further astrophysical information is used in defining the agnostic ensembles.}

The resulting ensemble forms the basis for the determination of the causal ceiling presented in the next section.

\section{Allowed EoS space at fixed $\Rof{1.4}$}

\begin{figure*}[t]
  \centering
  \includegraphics[width=0.88\textwidth]{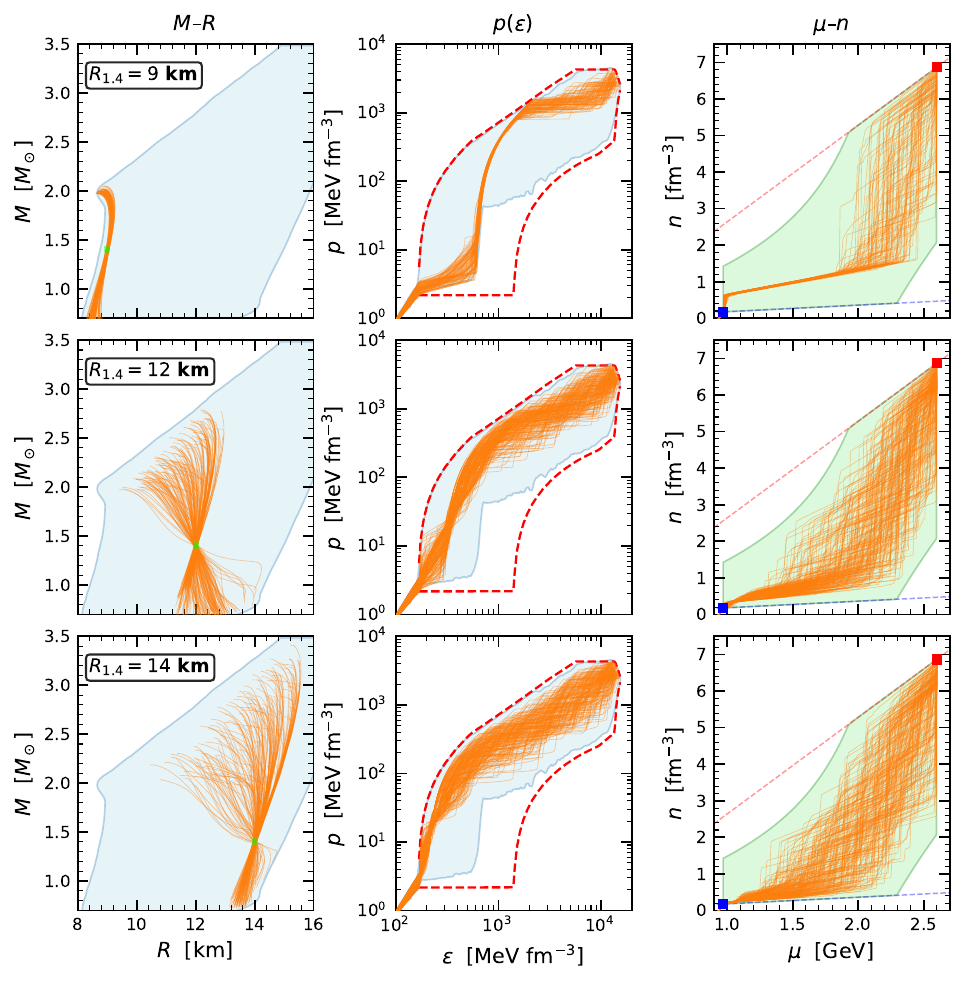}
  \caption{Fractal-Bridge 10-layer ensemble conditioned on three fixed small patches in the mass–radius plane: the radius at $1.4\Msun$, with $\R{1.4} = 9$, $12$, and $14$ km (rows from top to bottom). Columns show the mass--radius relation, the pressure $p(\varepsilon)$, and the chemical potential--density $\mu$--$n$ relation of the EoSs passing through the green patch in each $M$--$R$ panel (orange curves). The light-blue region in the $M$--$R$ and $p(\varepsilon)$ panels is the full fractal 10L domain; the light-green band in the $\mu$--$n$ panels is the outer constrained envelope~\cite{Komoltsev:2021jzg} of all stable, causal EoSs connecting the \chiEFT\ anchor (blue square) to the \pQCD\ anchor (red square). Red dashed curves in $p(\varepsilon)$ are the same allowed envelope of Ref.~\cite{Komoltsev:2021jzg} mapped to the $(\varepsilon, p)$ plane; blue and red dotted rays in $\mu$--$n$ are the $\cs=1$ rays through the \chiEFT\ and \pQCD\ anchors. Only $\MTOV > 1.97\Msun$ is imposed.}
  \label{fig:nine_panel}
\end{figure*}

Before identifying the causal ceiling, we first explore the space of EoSs compatible with a fixed value of $\Rof{1.4}$. Figure~\ref{fig:nine_panel} shows the subset of the fractal ensemble satisfying three representative values of $\Rof{1.4}$ together with the corresponding mass--radius relations, pressures $p(\varepsilon)$, and chemical potential--density relations $\mu$--$n$. Fixing $\Rof{1.4}$ already constrains both the EoS and the mass--radius relation far beyond the densities and masses directly probed by a $1.4\,M_\odot$ star, while the remaining freedom at higher densities depends strongly on the chosen value of $\Rof{1.4}$.

\emph{$\R{1.4}=9$~km (top row).} These are the softest low-density EoSs compatible with $\MTOV>1.97\Msun$. Satisfying the maximum-mass constraint forces them to stiffen almost immediately onto the $\cs=1$ trajectory, leaving essentially no freedom in the high-density continuation and producing a nearly unique post-$1.4\Msun$ mass--radius branch.

\emph{$\R{1.4}=12$~km (middle row).} The selected EoSs populate the interior of the allowed domain, leaving a moderate freedom in the high-density continuation and consequently a finite range of $\Rof{2.0}$ and $\MTOV$.

\emph{$\R{1.4}=14$~km (bottom row).} These correspond to stiff low-density EoSs lying close to the upper envelope at low density. Above $1.4\Msun$, however, they admit a broad range of high-density continuations---including trajectories with substantial softening or multiple phase transitions---while remaining compatible with $\MTOV>1.97\Msun$, resulting in the widest spread of $\Rof{2.0}$.

In the remainder of this work we focus exclusively on the upper envelope of this allowed space. For each fixed value of $\Rof{1.4}$, we identify the EoSs that maximize $\Rof{2.0}$ and thereby determine the causal ceiling $R_{2.0}^{\max}(R_{1.4})$.

\section{Causality bound}
\begin{figure*}[t]
  \centering
  \includegraphics[width=\textwidth]{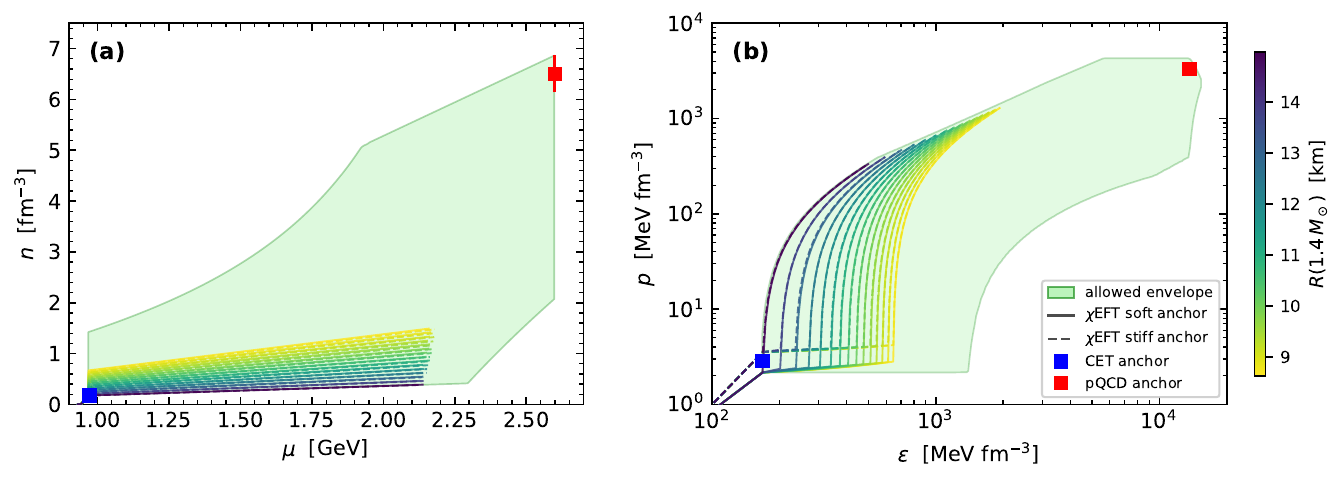}
  \caption{Origin of the bound. (a)~$\mu$--$n$ plane. The green band is the allowed envelope of thermodynamic consistency and causality between the \chiEFT\ (blue) and \pQCD\ (red) anchors~\cite{Komoltsev:2021jzg}. Thirty direct-causal EoSs are overlaid: solid for soft and dashed for stiff \chiEFT\ endpoint, colored by their resulting $\R{1.4}$. Each trajectory is truncated at the central density of its $\MTOV$ configuration. Over the stellar range the trajectories are straight rays through the \chiEFT\ anchor, $\mu(n)=(\mu_L/n_L)\,n$, as dictated by $\cs=1$. (b)~Same EoSs in the $\varepsilon$--$p$ plane. Along $\cs=1$ the relation is $p(\varepsilon) = \varepsilon + (2 p_L - \mu_L n_L)$, a slope-one line whose offset is fixed by the \chiEFT\ anchor. The bound of Fig.~\ref{fig:bound} is the stellar projection of these two elementary geometric statements.}
  \label{fig:geo}
\end{figure*}

\begin{figure*}[t]
  \centering
  \includegraphics[width=\textwidth]{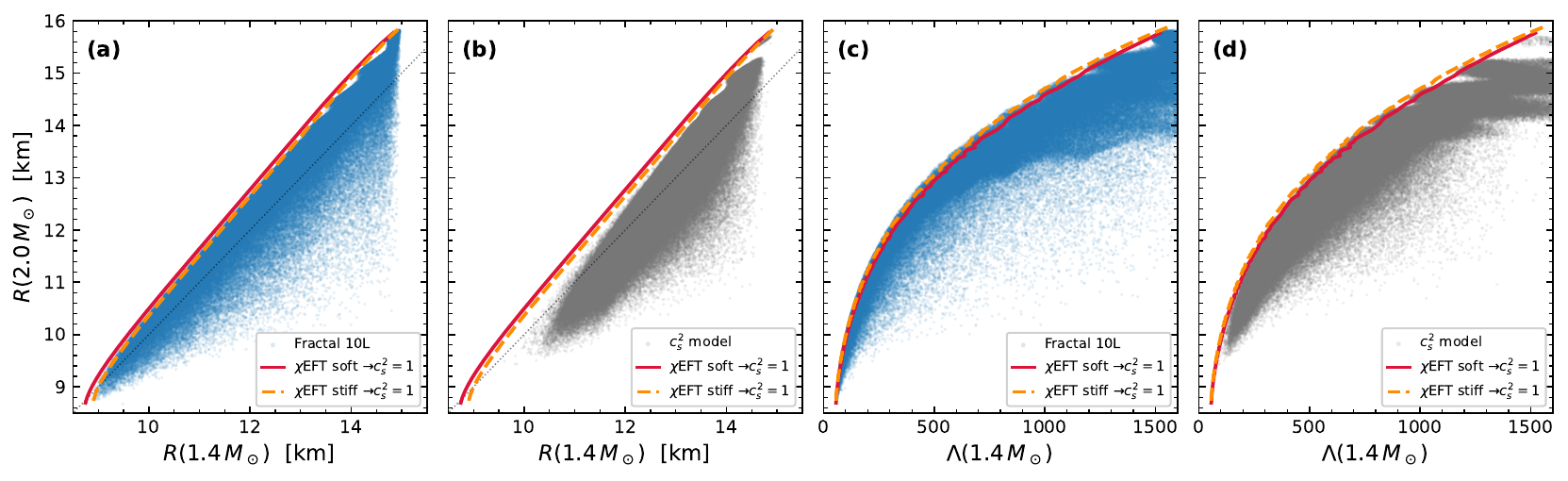}
  \caption{Causality bound on the neutron-star radius across masses. (a,b)~$\Rof{2.0}$ vs $\Rof{1.4}$ for the fractal-bridge 10-layer ensemble (blue) and the $\cs$ model (gray). (c,d)~$\Rof{2.0}$ vs $\Lambda_{1.4}$ for the same two ensembles. All four panels share the $\Rof{2.0}$ axis. In all panels the red solid and orange dashed curves are the \emph{direct-causal saturators}: EoSs in which $\cs=1$ is imposed immediately above the soft (solid) and stiff (dashed) \chiEFT\ endpoints. Both ensembles lie below this ceiling, but only the maximally agnostic fractal-bridge ensemble reaches it, whereas the less flexible $\cs(\varepsilon)$ ensemble falls short of the ceiling (see text). The dotted diagonal in (a,b) is $R_{2.0}=R_{1.4}$. Only $\MTOV>1.97\Msun$ is imposed.}
 \label{fig:bound}
\end{figure*}

Figure~\ref{fig:bound} reports the central result. Panels (a,b) show the distribution of $\R{1.4}$ and $\R{2.0}$ for each ensemble, and panels (c,d) the analogous $\R{2.0}$--$\Lambda_{1.4}$ correlation. The fractal cloud reaches further than the $\cs$ cloud at both small and large $\R{1.4}$, reflecting that the fractal prior samples a broader set of EoSs; the $\cs(\varepsilon)$ parametrization is only a subset of the fractal. {Quantitatively, at fixed $\R{1.4}=12\,$km the $68\%$ half-width of the fractal $\R{2.0}$ distribution exceeds the $\cs(\varepsilon)$ value by ${\simeq}25\%$ ($0.41$ vs $0.33\,$km). The difference is most pronounced near the upper envelope: the upper edge of the $\cs(\varepsilon)$ cloud falls short of the causal ceiling by ${\simeq}0.2\,$km at $\R{1.4}=12\,$km, growing to ${\simeq}0.5\,$km at $\R{1.4}=10\,$km, whereas the upper edge of the fractal cloud tracks the direct-causal saturators to within $0.05\,$km over the entire range of $\R{1.4}$.}

We identify the EoSs that saturate the upper envelope as a simple one-parameter family (the red and orange curves in Fig.~\ref{fig:bound}). These EoSs consist of a phase transition at $n=n_L$, followed by a maximally stiff segment with $c_s^2=1$ extending to the central density of a $1.97\,M_\odot$ star, depicted in Fig.~\ref{fig:geo}.

The strength of the phase transition parametrizes the family, which we refer to as the \emph{direct-causal saturators}. Propagating the uncertainty of the \chiEFT\ anchor at $n_L$ broadens the resulting line into the narrow red/orange band that traces the upper envelope in all four panels.

Quantitatively, the envelope is well described by a linear relation in $\R{1.4}$ and a power law in $\Lambda_{1.4}$,
\begin{align}
  \R{2.0}^{\max}(\R{1.4})      &\;=\; a_R\,\R{1.4} + b_R,           \label{eq:fit_RR} \\
  \R{2.0}^{\max}(\Lambda_{1.4})&\;=\; a_\Lambda\,\Lambda_{1.4}^{\,b_\Lambda},
                                                                     \label{eq:fit_RL}
\end{align}
with best-fit coefficients tabulated in Table~\ref{tab:saturator_fits}. Both \chiEFT\ soft and stiff endpoints yield nearly identical fits (relative deviation below $1\,\%$), and the residuals from the scan points are smaller than $0.08\,$km for all fits. Any EoS above these curves would require $\cs > 1$ somewhere in the star.

\begin{table}[t]
  \centering
  \caption{Fit coefficients of the direct-causal saturators after the $\MTOV > 1.97\Msun$ filter. Linear fit $\R{2.0} = a_R\,\R{1.4} + b_R$ and power law $\R{2.0} = a_\Lambda\,\Lambda_{1.4}^{\,b_\Lambda}$, with $\R{1.4}$ and $\R{2.0}$ in km.}
  \label{tab:saturator_fits}
  \begin{tabular}{lcccccc}
    \hline\hline
    \chiEFT\ anchor & $a_R$ & $b_R\,[\mathrm{km}]$ & rms [km] &
                      $a_\Lambda\,[\mathrm{km}]$ & $b_\Lambda$ & rms [km] \\
    \hline
    soft  & $1.155$ & $-1.109$ & $0.070$ & $4.474$ & $0.171$ & $0.055$ \\
    stiff & $1.158$ & $-1.264$ & $0.065$ & $4.500$ & $0.172$ & $0.059$ \\
    \hline\hline
  \end{tabular}
\end{table}

%

\section{Observational consequences}
\begin{figure*}[t]
  \centering
  \includegraphics[width=\textwidth]{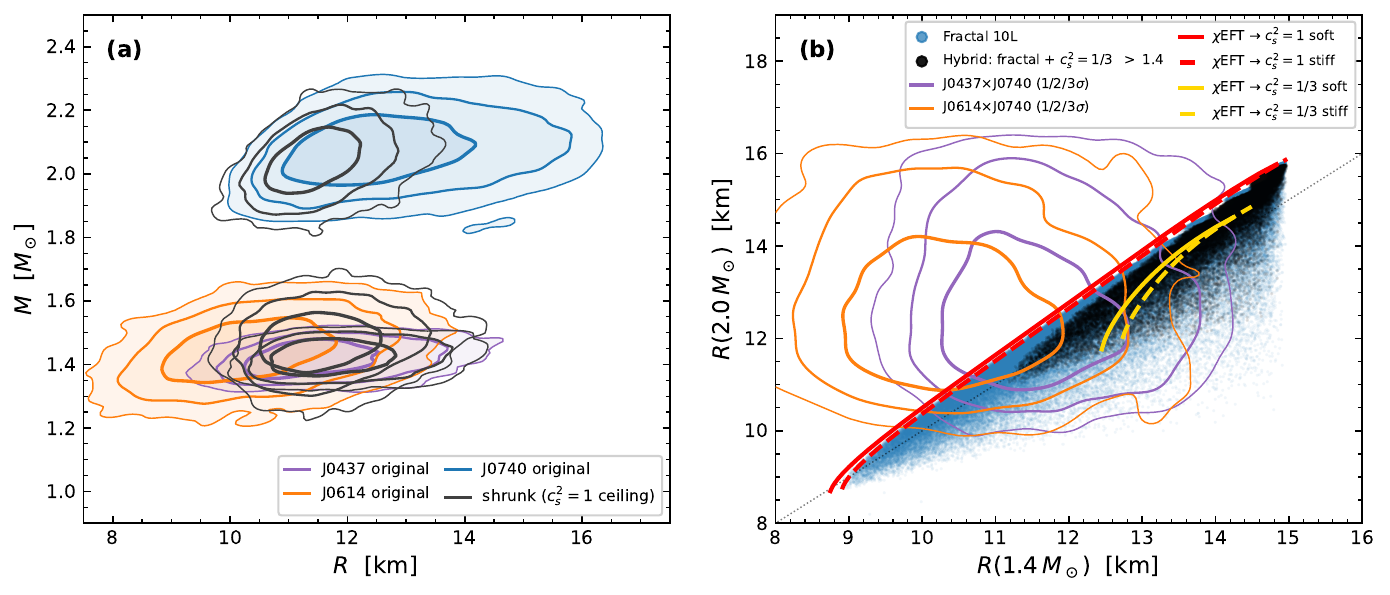}
  \caption{Two consequences of the causality ceiling for current data. (a)~$M$--$R$ plane. Filled purple/orange/blue contours are the original NICER $1/2/3\sigma$ posteriors of PSR~J0437$-$4715, PSR~J0614$-$3329 and PSR~J0740+6620, computed by Gaussian KDE on the published equal-weight posterior samples. Dark-gray overlaid contours ($1/2/3\sigma$) are the \emph{shrunk} posteriors obtained by re-sampling triples and retaining only those satisfying both $\R{2.0}\le a_R\,\R{1.4}+b_R$ pairings (Table~\ref{tab:saturator_fits}, \chiEFT\ soft). The shrink removes the large-$R$ tail of PSR~J0740 (median $\R{2.0}$ drops from $12.5$ to $\sim 11.4\,$km) and pushes PSR~J0437 and PSR~J0614 toward slightly larger radii. Only $7.5\%$ of the input triples survive. (b)~$\Rof{2.0}$--$\Rof{1.4}$ plane. Light-blue points: fully agnostic fractal-bridge 10L ensemble (any causal EoS above $1.4\Msun$). Dark-gray points: \emph{hybrid} family in which the fractal-bridge 10L EoS is kept up to the central energy density of the $1.4\Msun$ star, $\ecOf{1.4}$, and replaced by $\cs=1/3$ above; only $\MTOV>1.97\Msun$ EoSs are retained. Red solid/dashed and gold solid/dashed curves are the analytic \chiEFT$\to\cs=1$ and \chiEFT$\to\cs=1/3$ saturators (soft/stiff anchors). Purple and orange contours are the NICER joint posteriors J0437$\times$J0740 and J0614$\times$J0740 ($1/2/3\sigma$, independent product).}
  \label{fig:hybrid_shrunk}
\end{figure*}
The bound derived in Sec.~IV translates directly into two prior-free statements about current neutron-star observations. Both are visualized in Fig.~\ref{fig:hybrid_shrunk}.

\subsection{Refinement of independent radius measurements}
NICER provides statistically independent posterior distributions for PSR~J0437$-$4715, PSR~J0614$-$3329, and PSR~J0740+6620, with mean masses $\bar M=1.42\,\Msun$, $1.37\,\Msun$, and $2.07\,\Msun$, respectively. We approximate the two lighter stars as canonical $1.4\,\Msun$ stars and PSR~J0740+6620 as a canonical $2.0\,\Msun$ star, and denote the corresponding radius variables by $R_{0437}$, $R_{0614}$, and $R_{0740}$. This canonical-mass approximation is adopted only for definiteness; in Appendix~\ref{app:massexact} we repeat the conditioning with the causal ceiling evaluated at the exact mass of each sampled star and recover essentially the same radius posteriors; it does not affect any of our conclusions. The independent product posterior is then
\begin{equation}
P_0(R_{0437},R_{0614},R_{0740})
=
P(R_{0437})P(R_{0614})P(R_{0740}) .
\end{equation}
The causal ceiling defines the conditioned posterior
\begin{equation}
P_{\rm c}
=
\frac{P_0}{f_{\rm pass}}\,
\Theta( \R{2.0}^{\max}(R_{0437})-R_{0740})\,
\Theta( \R{2.0}^{\max}(R_{0614})-R_{0740})\,
\label{eq:constrained_posterior}
\end{equation}
where $f_{\rm pass}$ is the normalization constant of the conditioned posterior, equal to the fraction of the original joint posterior satisfying the causal constraint. Using the $\chi$EFT-soft fit in Table~\ref{tab:saturator_fits}, we obtain
\begin{equation}
f_{\rm pass}=7.5\%.
\end{equation}

The marginalized posterior distributions for the individual stars, obtained from the conditioned joint posterior $P_{\rm c}$, are shown as the dark-gray contours in Fig.~\ref{fig:hybrid_shrunk}(a). The refinement is driven primarily by the large-radius tail of PSR~J0740, whose posterior extends to $\R{2.0}\approx 16\,$km before conditioning but is truncated by the causal ceiling to $\R{2.0}\lesssim 12\,$km for the bulk of the surviving posterior mass. By contrast, the posterior distributions of the low-mass pulsars PSR~J0437 and PSR~J0614 shift only mildly toward larger radii, since the causal ceiling preferentially excludes combinations in which the low-mass stars occupy the small-radius tail of their distributions. 

We have verified that restricting the input samples to a narrow $\pm0.05\Msun$ window around the canonical masses before applying the conditioning yields essentially identical marginalized posteriors, implying that the canonical-mass approximation introduces a systematic uncertainty below the percent level.

\subsection{Lower bound on $\cs$ in dense cores}
The same data, viewed in the $(\R{1.4}, \R{2.0})$ plane, yield a \emph{lower} bound on the speed of sound deep inside the star. To make this manifest, we compare the NICER joint contours with three distinct EoS constructions that share the same low-density physics but differ in their high-density stiffness above the \chiEFT\ endpoint:
\begin{itemize}
   \item the \emph{fully agnostic} fractal-bridge 10-layer (10L) cloud (light blue), in which the EoS above $\ecOf{1.4}$ is unrestricted apart from thermodynamic stability and causality, encompassing every microphysical scenario from smooth nucleonic stiffening to strong first-order phase transitions, sound-speed peaks, and exotic-matter cores;
    
    \item the \emph{hybrid} cloud (black), obtained by truncating the fractal-bridge 10L EoS at $\ecOf{1.4}$ and continuing with constant $\cs=1/3$ above;
    
    \item the \emph{analytic conformal saturator} (yellow curves), the closed-form $\cs=1/3$ analog of the red $\cs=1$ ceiling: replace the entire density range above the \chiEFT\ endpoint by a constant $\cs=1/3$. Since $\cs=dp/d\varepsilon$, a segment of constant $\cs$ is a straight line of slope $\cs$ in the $\varepsilon$--$p$ plane; the conformal saturator is therefore a slope-$1/3$ line through the \chiEFT\ anchor, the analog of the slope-one $\cs=1$ ceiling [Fig.~\ref{fig:geo}(b)].
\end{itemize}

The hybrid cloud, which by construction populates the $\cs=1/3$-restricted region, lies systematically below the NICER joint contours: $\sim 90\%$ of the joint product posterior lies above the hybrid cloud at $1$--$2\sigma$. The analytic conformal saturator (yellow curves) makes the same statement in closed form and without any Monte-Carlo input. It is short---a constant $\cs=1/3$ above $n_L$ supports $\MTOV>1.97\Msun$ only over a narrow window of $\R{1.4}$, $\!\sim\!12.5$--$14\,$km---and it sits well below the bulk of the NICER joint posterior over its entire support. Outside this window the conformal-from-saturation hypothesis is already excluded by the $\MTOV$ requirement alone, so no $\cs=1/3$ EoS exists; inside it, the analytic saturator falsifies the hypothesis directly. Only the $\cs=1$ saturator (red curves) reaches the lower envelope of the NICER joint posterior.

The data therefore require that, somewhere above the \chiEFT\ endpoint $n_L$, the squared sound speed exceeds the conformal value: \emph{NICER plus causality}\;$\Rightarrow$\;\emph{$\cs > 1/3$ at some density above $n_L$}, with $\cs$ in fact close to unity for any joint NICER realization consistent with the ceiling. This conclusion is consistent with, and now follows independently from, the $\cs > 1/3$ statements of Refs.~\cite{Annala:2019puf,Altiparmak:2022bke,Annala:2023cwx}, which were drawn from parametric Bayesian inferences. Here it follows directly from the analytic ceiling, the analytic conformal saturator, and the raw NICER posteriors---without any EoS prior.

The two arguments are complementary. The first (NICER refinement, upper ceiling) is agnostic about $\cs$ within $[0,1]$ and just removes the unphysical upper tail in $\R{2.0}$; the second (the $\cs=1/3$ exclusion, lower envelope) is agnostic about the EoS family above the \chiEFT\ endpoint and isolates a specific stiffness constraint on the core. Neither uses any EoS prior beyond \chiEFT\ at saturation density and the causality limit.

\section{Discussion}
The radii (and tidal deformabilities) of different neutron stars are linked through the common underlying EoS. As a result, observations of one star constrain the properties inferred for other stars, even at different masses. This reduction in uncertainty, however, comes at the price of adopting an EoS prior. Here, we have quantified the consequences of assuming only that different neutron stars share a common EoS and that this EoS remains causal, with $c_s^2 \lesssim 1$.

These simple and general physical assumptions lead to a substantial reduction in the inferred uncertainties and, in particular, cleanly eliminate the large-radius tail of the PSR~J0740+6620 posterior. While this tail is known to be sensitive to the choice of prior in the inference, it is excluded here as a direct consequence of the physically well-motivated requirement of causality. The resulting limits thus define a robust null hypothesis, against which deviations induced by non-standard physics, such as dark-matter components or modifications of gravity, can be identified.

While even stronger constraints can be obtained by simultaneously inferring the EoS together with the masses and radii of multiple neutron stars \cite{Hoogkamer:2025eaq}, such analyses necessarily require specifying a particular construction for the EoS prior. In contrast, the procedure presented here assumes only that neutron stars share a common causal EoS, yet still yields comparable improvements in the inferred constraints, allowing the consequences of these generic physical assumptions to be disentangled from those associated with a particular choice of EoS prior. The present framework therefore isolates the impact of the common-EoS and causality assumptions, providing a transparent baseline against which more restrictive EoS constructions can be interpreted.

Importantly, the method operates entirely at the level of published posterior samples and therefore requires neither access to the original likelihood nor a reanalysis of the observational data. It thus provides a simple and practical framework for rapidly assessing the impact of alternative physical scenarios without the need to repeat a full hierarchical inference.

Although we have focused on current NICER observations, the method is immediately applicable to any collection of independent radius or tidal-deformability posteriors and will become increasingly powerful as additional measurements become available.

The limits presented here depend only mildly on the low-density EoS obtained from $\chi$EFT, making them robust against uncertainties in the current calculation, although significant improvements in the $\chi$EFT input would lead to a moderate tightening of the bounds. In contrast, the extremal EoSs that determine these limits are insensitive to the constraints imposed by perturbative QCD at high densities.

On the technical side, we find that the recent fractal-bridge prior of \cite{Gorda:2025aiu} explores the EoS phase space significantly more broadly than the commonly used $c_s^2$ prior \cite{Annala:2017llu}. This makes it well suited for establishing robust bounds under minimal assumptions and enables the identification of extremal configurations.

\begin{acknowledgments}
The authors are listed alphabetically. 
 A.K. was supported by the Research Council of Norway through the FRIPRO programme (CoreQCD, project number 361873) and in part by grant NSF PHY-2309135 to the Kavli Institute for Theoretical Physics (KITP). T.M.\ would like to acknowledge the support of FCT (Funda\c{c}\~{a}o para a Ci\^{e}ncia e a Tecnologia, I.P., Portugal) under project UID/04564/2025, identified by DOI 10.54499/UIDB/04564/2025.
T.M.\ would also like to thank the FCT Mobility grant (Reference: FCT/Mobility/1381803446/2024-25) and the University of Stavanger, Department of Mathematics and Physics, for the local hospitality during the visit for this work.
\end{acknowledgments}

\appendix
\section{Robustness of the posterior conditioning against the fixed canonical-mass approximation}
\label{app:massexact}
{The refinement of independent NICER measured posteriors in Sec.~V fixes the three pulsars at the canonical reference masses: the two lighter stars at $1.4\Msun$ and PSR~J0740+6620 at $2.0\Msun$, close to their respective posterior means, so that the sampled radii are identified with $\R{1.4}$ and $\R{2.0}$, and the ceiling is applied through the fixed linear relation of Table~\ref{tab:saturator_fits}. The NICER posteriors, however, carry substantial mass uncertainties---the equal-weight samples span $1.29$--$1.53\Msun$ (J0437), $1.16$--$1.73\Msun$ (J0614) and $1.79$--$2.38\Msun$ (J0740). Here we verify that this approximation does not bias the result.} 

\begin{figure*}[t]
  \centering
  \includegraphics[width=\textwidth]{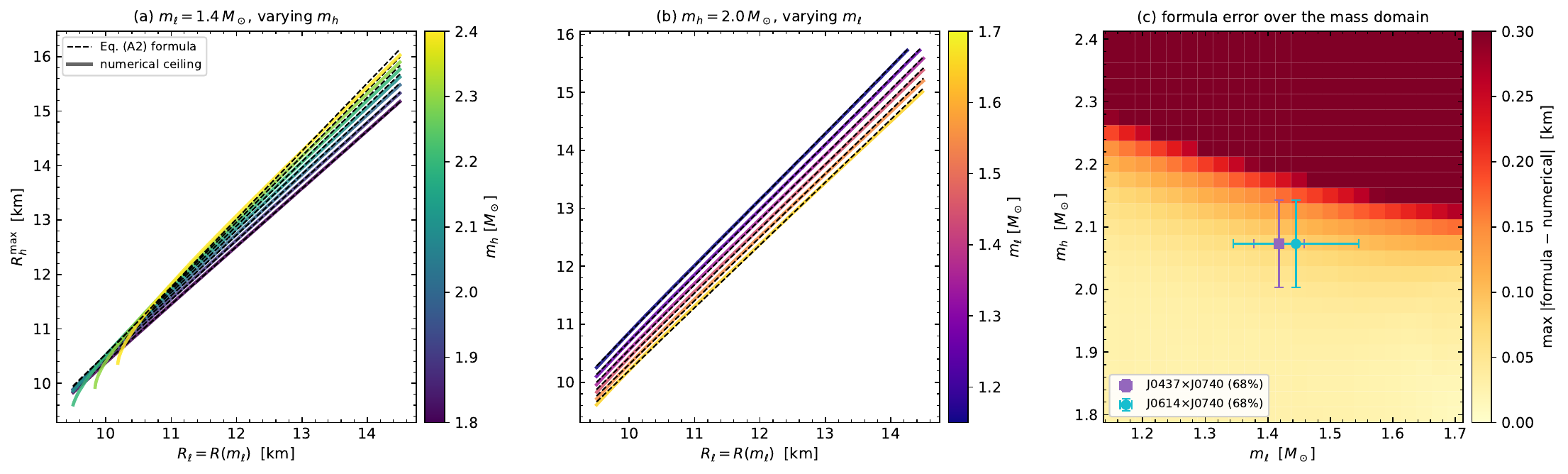}
  \caption{Accuracy of the closed-form exact-mass ceiling, Eq.~(\ref{eq:massexact_surface}) with the coefficients of Table~\ref{tab:massexact_fit}. (a)~Numerical family ceilings (solid, colored by $m_h$) and the closed form (black dashed) at fixed $m_\ell = 1.4\Msun$. (b)~Same at fixed $m_h = 2.0\Msun$, varying $m_\ell$ (colored). The numerical ceilings are smooth across the full radius range, and the closed form tracks them throughout, deviating only toward the extreme low-$R_\ell$ and high-$m_h$ corners. (c)~Maximum absolute deviation between the closed form and the numerical ceilings over $9.5 \le R_\ell \le 14.5\,$km, as a function of the mass pair. The markers show the median masses and $68\%$ intervals of the J0437$\times$J0740 and J0614$\times$J0740 NICER pairings: within their $68\%$ mass ranges the closed form is accurate to better than $0.24\,$km, and to $\lesssim 0.1\,$km at the medians.} 
  \label{fig:surfacefit}
\end{figure*}

\begin{figure}[!tb]
  \centering
  \includegraphics[width=\columnwidth]{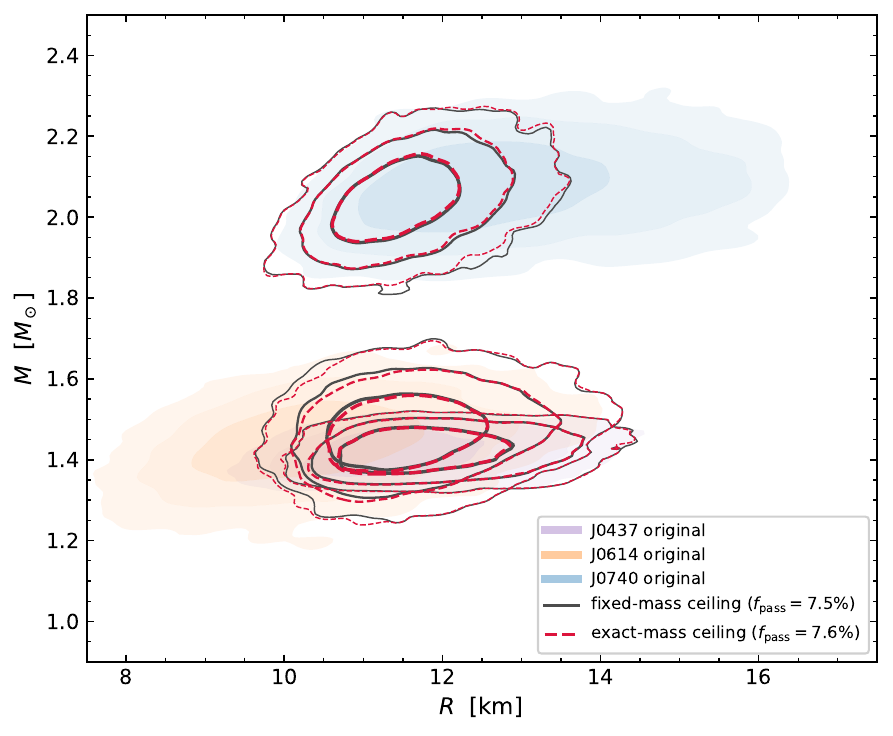}
  \caption{{Conditioned NICER posteriors from the fixed-mass and exact-mass causal ceilings. Filled contours: original $1/2/3\sigma$ posteriors of PSR~J0437$-$4715 (purple), PSR~J0614$-$3329 (orange) and PSR~J0740+6620 (blue). Solid dark-gray contours: posteriors conditioned with the fixed-mass ceiling of Eq.~(\ref{eq:constrained_posterior}), evaluated at the canonical masses $(1.4,\,2.0)\Msun$ (see Table~\ref{tab:saturator_fits}, \chiEFT\ soft), as in Fig.~\ref{fig:hybrid_shrunk}(a). Dashed red contours: posteriors conditioned with the exact-mass ceiling $R^{\max}_{(m_\ell,\,m_h)}$ reconstructed for the sampled mass pair of every posterior triple. The two conditionings retain $7.5\%$ and $7.6\%$ of the joint posterior, respectively, and produce statistically indistinguishable marginals (per-pulsar Kolmogorov--Smirnov distance $\le 0.02$ \cite{Massey01031951}; credible radius bounds agreeing to within $0.04\,$km).}} \label{fig:massexact}
\end{figure}

{Because the direct-causal saturators form a one-parameter family---the strength of the phase transition at $n_L$, followed by the $\cs=1$ segment---the causal ceiling is available for \emph{any} mass pair, not only for $(1.4,\,2.0)\Msun$: for masses $(m_\ell, m_h)$ the family traces the parametric curve $\{(R_j(m_\ell),\,R_j(m_h))\}_j$ in the radius--radius plane, of which the relation of Table~\ref{tab:saturator_fits} is the $(1.4,\,2.0)\Msun$ member. We therefore repeat the conditioning of Eq.~(\ref{eq:constrained_posterior}) with the fixed relation replaced, sample by sample, by the exact-mass constraint
\begin{equation}
  R_h \;\le\; R^{\max}_{(m_\ell,\,m_h)}(R_\ell),
  \label{eq:massexact_cut}
\end{equation}
imposed for both pairings (J0437, J0740) and (J0614, J0740), where $(m_\ell, R_\ell)$ and $(m_h, R_h)$ are the masses and radii actually drawn from the respective posteriors.}

{Numerically, the family is realized by a bank of $1.5\times10^4$ \chiEFT-soft saturator EoSs spanning phase-transition strengths from $n_{\rm jump}=n_L$ (no transition, stiffest) to the softest configuration still reaching $\MTOV>1.97\Msun$. For each EoS the full $M(\varepsilon_c)$, $R(\varepsilon_c)$ sequence is solved on a dense central-density grid, refined around the maximum-mass configuration, and the stable branch is tabulated as $R_j(M)$ on a $0.005\Msun$ grid; the exact-mass ceiling is the upper envelope of $\{(R_j(m_\ell),R_j(m_h))\}_j$.}

{The exact-mass ceilings also admit a simple closed-form approximation. Over the observationally relevant domain $1.15 \le m_\ell \le 1.70\Msun$, $1.80 \le m_h \le 2.40\Msun$ and $9.5 \le R_\ell \le 14.5\,$km, each fixed-mass-pair ceiling is closely linear in $R_\ell$ (rms $\le 0.08\,$km, median $0.03\,$km), and the fitted slope and intercept vary smoothly with the two masses, so that
\begin{equation}
  R^{\max}_{(m_\ell,\,m_h)}(R_\ell) \;\simeq\;
  a_R(m_\ell, m_h)\, R_\ell \;+\; b_R(m_\ell, m_h),
  \label{eq:massexact_surface}
\end{equation}
with $a_R$ and $b_R$ quadratic polynomials in the masses whose coefficients are listed in Table~\ref{tab:massexact_fit}. The quality of this closed form is shown in Fig.~\ref{fig:surfacefit}: panels (a,b) overlay Eq.~(\ref{eq:massexact_surface}) on the numerically constructed ceilings for representative mass cuts, and panel (c) maps its maximum deviation over the mass domain. At $(m_\ell, m_h) = (1.4,\,2.0)\Msun$, Eq.~(\ref{eq:massexact_surface}) reproduces the Table~\ref{tab:saturator_fits} relation to better than $0.12\,$km across the fit window. Against the numerically constructed ceilings, the closed form has an rms error of $0.06\,$km, with $95\%$ of evaluations within $0.08\,$km; the largest single-point deviations ($\lesssim 0.45\,$km) are confined to the extreme corners of the domain---the strong-transition end $R_\ell \simeq 9.5\,$km and the heaviest $m_h \gtrsim 2.35\Msun$---where the ceiling is intrinsically curved and which lie far in the tails of the NICER posteriors (the J0740 posterior places $<1\%$ of its mass above $2.25\Msun$). Within the $68\%$ mass ranges of the two pulsar pairings the closed form is accurate to better than $0.24\,$km (Fig.~\ref{fig:surfacefit}(c)). All quantitative results quoted in this appendix use the numerical ceilings; replacing them by Eq.~(\ref{eq:massexact_surface}) changes the surviving fraction from $7.6\%$ to $7.3\%$.}

\begin{table}[H]
  \centering
  \caption{Coefficients of the quadratic mass dependence of the exact-mass
  ceiling, Eq.~(\ref{eq:massexact_surface}):
  $a_R$ or $b_R = c_0 + c_1 m_\ell + c_2 m_h + c_3 m_\ell m_h
  + c_4 m_\ell^2 + c_5 m_h^2$, with masses in $M_\odot$ and $b_R$ in km.
  Valid for $1.15 \le m_\ell \le 1.70$, $1.80 \le m_h \le 2.40$,
  $9.5 \le R_\ell \le 14.5\,$km (\chiEFT-soft anchor).}
  \label{tab:massexact_fit}
  \begin{tabular}{lcccccc}
    \hline\hline
     & $c_0$ & $c_1$ & $c_2$ & $c_3$ & $c_4$ & $c_5$ \\
    \hline
    $a_R$            & $1.143$  & $-0.122$ & $-0.141$ & $0.022$  & $-0.019$ & $0.100$  \\
    $b_R$ [km]       & $-2.901$ & $-0.835$ & $5.356$  & $-0.644$ & $0.760$  & $-1.748$ \\
    \hline\hline
  \end{tabular}
\end{table}

The result is shown in Fig.~\ref{fig:massexact}. Replacing the fixed-mass ceiling of the main text by the exact-mass ceiling changes the retained fraction from $7.5\%$ to $7.6\%$, with the two conditioned posteriors visually indistinguishable. Quantitatively, the median and the $68\%$ and $95\%$ credible radius bounds of all three pulsars agree between the two methods to within $0.04\,$km, and the corresponding shrunk marginals differ by a Kolmogorov--Smirnov distance $\le 0.02$ \cite{Massey01031951} ---i.e.\ the two posteriors are statistically indistinguishable.


\begin{thebibliography}{28}%
\makeatletter
\providecommand \@ifxundefined [1]{%
 \@ifx{#1\undefined}
}%
\providecommand \@ifnum [1]{%
 \ifnum #1\expandafter \@firstoftwo
 \else \expandafter \@secondoftwo
 \fi
}%
\providecommand \@ifx [1]{%
 \ifx #1\expandafter \@firstoftwo
 \else \expandafter \@secondoftwo
 \fi
}%
\providecommand \natexlab [1]{#1}%
\providecommand \enquote  [1]{``#1''}%
\providecommand \bibnamefont  [1]{#1}%
\providecommand \bibfnamefont [1]{#1}%
\providecommand \citenamefont [1]{#1}%
\providecommand \href@noop [0]{\@secondoftwo}%
\providecommand \href [0]{\begingroup \@sanitize@url \@href}%
\providecommand \@href[1]{\@@startlink{#1}\@@href}%
\providecommand \@@href[1]{\endgroup#1\@@endlink}%
\providecommand \@sanitize@url [0]{\catcode `\\12\catcode `\$12\catcode
  `\&12\catcode `\#12\catcode `\^12\catcode `\_12\catcode `\%12\relax}%
\providecommand \@@startlink[1]{}%
\providecommand \@@endlink[0]{}%
\providecommand \url  [0]{\begingroup\@sanitize@url \@url }%
\providecommand \@url [1]{\endgroup\@href {#1}{\urlprefix }}%
\providecommand \urlprefix  [0]{URL }%
\providecommand \Eprint [0]{\href }%
\providecommand \doibase [0]{https://doi.org/}%
\providecommand \selectlanguage [0]{\@gobble}%
\providecommand \bibinfo  [0]{\@secondoftwo}%
\providecommand \bibfield  [0]{\@secondoftwo}%
\providecommand \translation [1]{[#1]}%
\providecommand \BibitemOpen [0]{}%
\providecommand \bibitemStop [0]{}%
\providecommand \bibitemNoStop [0]{.\EOS\space}%
\providecommand \EOS [0]{\spacefactor3000\relax}%
\providecommand \BibitemShut  [1]{\csname bibitem#1\endcsname}%
\let\auto@bib@innerbib\@empty
\bibitem [{\citenamefont {Cromartie}\ \emph {et~al.}(2019)\citenamefont
  {Cromartie} \emph {et~al.}}]{NANOGrav:2019jur}%
  \BibitemOpen
  \bibfield  {author} {\bibinfo {author} {\bibfnamefont {H.~T.}\ \bibnamefont
  {Cromartie}} \emph {et~al.} (\bibinfo {collaboration} {NANOGrav}),\ }\href
  {https://doi.org/10.1038/s41550-019-0880-2} {\bibfield  {journal} {\bibinfo
  {journal} {Nature Astron.}\ }\textbf {\bibinfo {volume} {4}},\ \bibinfo
  {pages} {72} (\bibinfo {year} {2019})},\ \Eprint
  {https://arxiv.org/abs/1904.06759} {arXiv:1904.06759 [astro-ph.HE]}
  \BibitemShut {NoStop}%
\bibitem [{\citenamefont {Fonseca}\ \emph {et~al.}(2021)\citenamefont {Fonseca}
  \emph {et~al.}}]{Fonseca:2021wxt}%
  \BibitemOpen
  \bibfield  {author} {\bibinfo {author} {\bibfnamefont {E.}~\bibnamefont
  {Fonseca}} \emph {et~al.},\ }\href {https://doi.org/10.3847/2041-8213/ac03b8}
  {\bibfield  {journal} {\bibinfo  {journal} {Astrophys. J. Lett.}\ }\textbf
  {\bibinfo {volume} {915}},\ \bibinfo {pages} {L12} (\bibinfo {year}
  {2021})},\ \Eprint {https://arxiv.org/abs/2104.00880} {arXiv:2104.00880
  [astro-ph.HE]} \BibitemShut {NoStop}%
\bibitem [{\citenamefont {Antoniadis}\ \emph {et~al.}(2013)\citenamefont
  {Antoniadis} \emph {et~al.}}]{Antoniadis:2013pzd}%
  \BibitemOpen
  \bibfield  {author} {\bibinfo {author} {\bibfnamefont {J.}~\bibnamefont
  {Antoniadis}} \emph {et~al.},\ }\href
  {https://doi.org/10.1126/science.1233232} {\bibfield  {journal} {\bibinfo
  {journal} {Science}\ }\textbf {\bibinfo {volume} {340}},\ \bibinfo {pages}
  {6131} (\bibinfo {year} {2013})},\ \Eprint {https://arxiv.org/abs/1304.6875}
  {arXiv:1304.6875 [astro-ph.HE]} \BibitemShut {NoStop}%
\bibitem [{\citenamefont {Abbott}\ \emph {et~al.}(2018)\citenamefont {Abbott}
  \emph {et~al.}}]{LIGOScientific:2018cki}%
  \BibitemOpen
  \bibfield  {author} {\bibinfo {author} {\bibfnamefont {B.~P.}\ \bibnamefont
  {Abbott}} \emph {et~al.} (\bibinfo {collaboration} {LIGO Scientific,
  Virgo}),\ }\href {https://doi.org/10.1103/PhysRevLett.121.161101} {\bibfield
  {journal} {\bibinfo  {journal} {Phys. Rev. Lett.}\ }\textbf {\bibinfo
  {volume} {121}},\ \bibinfo {pages} {161101} (\bibinfo {year} {2018})},\
  \Eprint {https://arxiv.org/abs/1805.11581} {arXiv:1805.11581 [gr-qc]}
  \BibitemShut {NoStop}%
\bibitem [{\citenamefont {Miller}\ \emph {et~al.}(2019)\citenamefont {Miller}
  \emph {et~al.}}]{Miller:2019cac}%
  \BibitemOpen
  \bibfield  {author} {\bibinfo {author} {\bibfnamefont {M.~C.}\ \bibnamefont
  {Miller}} \emph {et~al.},\ }\href {https://doi.org/10.3847/2041-8213/ab50c5}
  {\bibfield  {journal} {\bibinfo  {journal} {Astrophys. J. Lett.}\ }\textbf
  {\bibinfo {volume} {887}},\ \bibinfo {pages} {L24} (\bibinfo {year}
  {2019})},\ \Eprint {https://arxiv.org/abs/1912.05705} {arXiv:1912.05705
  [astro-ph.HE]} \BibitemShut {NoStop}%
\bibitem [{\citenamefont {Riley}\ \emph {et~al.}(2019)\citenamefont {Riley}
  \emph {et~al.}}]{Riley:2019yda}%
  \BibitemOpen
  \bibfield  {author} {\bibinfo {author} {\bibfnamefont {T.~E.}\ \bibnamefont
  {Riley}} \emph {et~al.},\ }\href {https://doi.org/10.3847/2041-8213/ab481c}
  {\bibfield  {journal} {\bibinfo  {journal} {Astrophys. J. Lett.}\ }\textbf
  {\bibinfo {volume} {887}},\ \bibinfo {pages} {L21} (\bibinfo {year}
  {2019})},\ \Eprint {https://arxiv.org/abs/1912.05702} {arXiv:1912.05702
  [astro-ph.HE]} \BibitemShut {NoStop}%
\bibitem [{\citenamefont {Vinciguerra}\ \emph {et~al.}(2024)\citenamefont
  {Vinciguerra} \emph {et~al.}}]{Vinciguerra:2023qxq}%
  \BibitemOpen
  \bibfield  {author} {\bibinfo {author} {\bibfnamefont {S.}~\bibnamefont
  {Vinciguerra}} \emph {et~al.},\ }\href
  {https://doi.org/10.3847/1538-4357/acfb83} {\bibfield  {journal} {\bibinfo
  {journal} {Astrophys. J.}\ }\textbf {\bibinfo {volume} {961}},\ \bibinfo
  {pages} {62} (\bibinfo {year} {2024})},\ \Eprint
  {https://arxiv.org/abs/2308.09469} {arXiv:2308.09469 [astro-ph.HE]}
  \BibitemShut {NoStop}%
\bibitem [{\citenamefont {Miller}\ \emph {et~al.}(2021)\citenamefont {Miller}
  \emph {et~al.}}]{Miller:2021qha}%
  \BibitemOpen
  \bibfield  {author} {\bibinfo {author} {\bibfnamefont {M.~C.}\ \bibnamefont
  {Miller}} \emph {et~al.},\ }\href {https://doi.org/10.3847/2041-8213/ac089b}
  {\bibfield  {journal} {\bibinfo  {journal} {Astrophys. J. Lett.}\ }\textbf
  {\bibinfo {volume} {918}},\ \bibinfo {pages} {L28} (\bibinfo {year}
  {2021})},\ \Eprint {https://arxiv.org/abs/2105.06979} {arXiv:2105.06979
  [astro-ph.HE]} \BibitemShut {NoStop}%
\bibitem [{\citenamefont {Riley}\ \emph {et~al.}(2021)\citenamefont {Riley}
  \emph {et~al.}}]{Riley:2021pdl}%
  \BibitemOpen
  \bibfield  {author} {\bibinfo {author} {\bibfnamefont {T.~E.}\ \bibnamefont
  {Riley}} \emph {et~al.},\ }\href {https://doi.org/10.3847/2041-8213/ac0a81}
  {\bibfield  {journal} {\bibinfo  {journal} {Astrophys. J. Lett.}\ }\textbf
  {\bibinfo {volume} {918}},\ \bibinfo {pages} {L27} (\bibinfo {year}
  {2021})},\ \Eprint {https://arxiv.org/abs/2105.06980} {arXiv:2105.06980
  [astro-ph.HE]} \BibitemShut {NoStop}%
\bibitem [{\citenamefont {Salmi}\ \emph {et~al.}(2024)\citenamefont {Salmi}
  \emph {et~al.}}]{Salmi:2024aum}%
  \BibitemOpen
  \bibfield  {author} {\bibinfo {author} {\bibfnamefont {T.}~\bibnamefont
  {Salmi}} \emph {et~al.},\ }\href {https://doi.org/10.3847/1538-4357/ad5f1f}
  {\bibfield  {journal} {\bibinfo  {journal} {Astrophys. J.}\ }\textbf
  {\bibinfo {volume} {974}},\ \bibinfo {pages} {294} (\bibinfo {year}
  {2024})},\ \Eprint {https://arxiv.org/abs/2406.14466} {arXiv:2406.14466
  [astro-ph.HE]} \BibitemShut {NoStop}%
\bibitem [{\citenamefont {Dittmann}\ \emph {et~al.}(2024)\citenamefont
  {Dittmann} \emph {et~al.}}]{Dittmann:2024mbo}%
  \BibitemOpen
  \bibfield  {author} {\bibinfo {author} {\bibfnamefont {A.~J.}\ \bibnamefont
  {Dittmann}} \emph {et~al.},\ }\href
  {https://doi.org/10.3847/1538-4357/ad5f1e} {\bibfield  {journal} {\bibinfo
  {journal} {Astrophys. J.}\ }\textbf {\bibinfo {volume} {974}},\ \bibinfo
  {pages} {295} (\bibinfo {year} {2024})},\ \Eprint
  {https://arxiv.org/abs/2406.14467} {arXiv:2406.14467 [astro-ph.HE]}
  \BibitemShut {NoStop}%
\bibitem [{\citenamefont {Choudhury}\ \emph {et~al.}(2024)\citenamefont
  {Choudhury} \emph {et~al.}}]{Choudhury:2024xbk}%
  \BibitemOpen
  \bibfield  {author} {\bibinfo {author} {\bibfnamefont {D.}~\bibnamefont
  {Choudhury}} \emph {et~al.},\ }\href
  {https://doi.org/10.3847/2041-8213/ad5a6f} {\bibfield  {journal} {\bibinfo
  {journal} {Astrophys. J. Lett.}\ }\textbf {\bibinfo {volume} {971}},\
  \bibinfo {pages} {L20} (\bibinfo {year} {2024})},\ \Eprint
  {https://arxiv.org/abs/2407.06789} {arXiv:2407.06789 [astro-ph.HE]}
  \BibitemShut {NoStop}%
\bibitem [{\citenamefont {Mauviard}\ \emph {et~al.}(2025)\citenamefont
  {Mauviard} \emph {et~al.}}]{Mauviard:2025dmd}%
  \BibitemOpen
  \bibfield  {author} {\bibinfo {author} {\bibfnamefont {L.}~\bibnamefont
  {Mauviard}} \emph {et~al.},\ }\href
  {https://doi.org/10.3847/1538-4357/ae145d} {\bibfield  {journal} {\bibinfo
  {journal} {Astrophys. J.}\ }\textbf {\bibinfo {volume} {995}},\ \bibinfo
  {pages} {60} (\bibinfo {year} {2025})},\ \Eprint
  {https://arxiv.org/abs/2506.14883} {arXiv:2506.14883 [astro-ph.HE]}
  \BibitemShut {NoStop}%
\bibitem [{\citenamefont {Drischler}\ \emph {et~al.}(2021)\citenamefont
  {Drischler}, \citenamefont {Han}, \citenamefont {Lattimer}, \citenamefont
  {Prakash}, \citenamefont {Reddy},\ and\ \citenamefont
  {Zhao}}]{Drischler:2020fvz}%
  \BibitemOpen
  \bibfield  {author} {\bibinfo {author} {\bibfnamefont {C.}~\bibnamefont
  {Drischler}}, \bibinfo {author} {\bibfnamefont {S.}~\bibnamefont {Han}},
  \bibinfo {author} {\bibfnamefont {J.~M.}\ \bibnamefont {Lattimer}}, \bibinfo
  {author} {\bibfnamefont {M.}~\bibnamefont {Prakash}}, \bibinfo {author}
  {\bibfnamefont {S.}~\bibnamefont {Reddy}},\ and\ \bibinfo {author}
  {\bibfnamefont {T.}~\bibnamefont {Zhao}},\ }\href
  {https://doi.org/10.1103/PhysRevC.103.045808} {\bibfield  {journal} {\bibinfo
   {journal} {Phys. Rev. C}\ }\textbf {\bibinfo {volume} {103}},\ \bibinfo
  {pages} {045808} (\bibinfo {year} {2021})},\ \Eprint
  {https://arxiv.org/abs/2009.06441} {arXiv:2009.06441 [nucl-th]} \BibitemShut
  {NoStop}%
\bibitem [{\citenamefont {Lin}\ and\ \citenamefont
  {Steiner}(2024)}]{Lin:2023cbo}%
  \BibitemOpen
  \bibfield  {author} {\bibinfo {author} {\bibfnamefont {Z.}~\bibnamefont
  {Lin}}\ and\ \bibinfo {author} {\bibfnamefont {A.~W.}\ \bibnamefont
  {Steiner}},\ }\href {https://doi.org/10.3847/2041-8213/ad7eb5} {\bibfield
  {journal} {\bibinfo  {journal} {Astrophys. J. Lett.}\ }\textbf {\bibinfo
  {volume} {974}},\ \bibinfo {pages} {L17} (\bibinfo {year} {2024})},\ \Eprint
  {https://arxiv.org/abs/2310.01619} {arXiv:2310.01619 [astro-ph.HE]}
  \BibitemShut {NoStop}%
\bibitem [{\citenamefont {Tang}\ \emph {et~al.}(2025)\citenamefont {Tang},
  \citenamefont {Huang},\ and\ \citenamefont {Fan}}]{Tang:2025xib}%
  \BibitemOpen
  \bibfield  {author} {\bibinfo {author} {\bibfnamefont {S.-P.}\ \bibnamefont
  {Tang}}, \bibinfo {author} {\bibfnamefont {Y.-J.}\ \bibnamefont {Huang}},\
  and\ \bibinfo {author} {\bibfnamefont {Y.-Z.}\ \bibnamefont {Fan}},\ }\href
  {https://doi.org/10.1103/bmsk-8n85} {\bibfield  {journal} {\bibinfo
  {journal} {Phys. Rev. D}\ }\textbf {\bibinfo {volume} {112}},\ \bibinfo
  {pages} {083009} (\bibinfo {year} {2025})},\ \Eprint
  {https://arxiv.org/abs/2507.10025} {arXiv:2507.10025 [astro-ph.HE]}
  \BibitemShut {NoStop}%
\bibitem [{\citenamefont {Ferreira}\ and\ \citenamefont
  {Provid{\^e}ncia}(2024)}]{Ferreira:2024hxc}%
  \BibitemOpen
  \bibfield  {author} {\bibinfo {author} {\bibfnamefont {M.}~\bibnamefont
  {Ferreira}}\ and\ \bibinfo {author} {\bibfnamefont {C.}~\bibnamefont
  {Provid{\^e}ncia}},\ }\href {https://doi.org/10.1103/PhysRevD.110.063018}
  {\bibfield  {journal} {\bibinfo  {journal} {Phys. Rev. D}\ }\textbf {\bibinfo
  {volume} {110}},\ \bibinfo {pages} {063018} (\bibinfo {year} {2024})},\
  \Eprint {https://arxiv.org/abs/2406.12582} {arXiv:2406.12582 [nucl-th]}
  \BibitemShut {NoStop}%
\bibitem [{\citenamefont {Annala}\ \emph {et~al.}(2020)\citenamefont {Annala},
  \citenamefont {Gorda}, \citenamefont {Kurkela}, \citenamefont
  {N{\"a}ttil{\"a}},\ and\ \citenamefont {Vuorinen}}]{Annala:2019puf}%
  \BibitemOpen
  \bibfield  {author} {\bibinfo {author} {\bibfnamefont {E.}~\bibnamefont
  {Annala}}, \bibinfo {author} {\bibfnamefont {T.}~\bibnamefont {Gorda}},
  \bibinfo {author} {\bibfnamefont {A.}~\bibnamefont {Kurkela}}, \bibinfo
  {author} {\bibfnamefont {J.}~\bibnamefont {N{\"a}ttil{\"a}}},\ and\ \bibinfo
  {author} {\bibfnamefont {A.}~\bibnamefont {Vuorinen}},\ }\href
  {https://doi.org/10.1038/s41567-020-0914-9} {\bibfield  {journal} {\bibinfo
  {journal} {Nature Phys.}\ }\textbf {\bibinfo {volume} {16}},\ \bibinfo
  {pages} {907} (\bibinfo {year} {2020})},\ \Eprint
  {https://arxiv.org/abs/1903.09121} {arXiv:1903.09121 [astro-ph.HE]}
  \BibitemShut {NoStop}%
\bibitem [{\citenamefont {Annala}\ \emph {et~al.}(2023)\citenamefont {Annala},
  \citenamefont {Gorda}, \citenamefont {Hirvonen}, \citenamefont {Komoltsev},
  \citenamefont {Kurkela}, \citenamefont {N{\"a}ttil{\"a}},\ and\ \citenamefont
  {Vuorinen}}]{Annala:2023cwx}%
  \BibitemOpen
  \bibfield  {author} {\bibinfo {author} {\bibfnamefont {E.}~\bibnamefont
  {Annala}}, \bibinfo {author} {\bibfnamefont {T.}~\bibnamefont {Gorda}},
  \bibinfo {author} {\bibfnamefont {J.}~\bibnamefont {Hirvonen}}, \bibinfo
  {author} {\bibfnamefont {O.}~\bibnamefont {Komoltsev}}, \bibinfo {author}
  {\bibfnamefont {A.}~\bibnamefont {Kurkela}}, \bibinfo {author} {\bibfnamefont
  {J.}~\bibnamefont {N{\"a}ttil{\"a}}},\ and\ \bibinfo {author} {\bibfnamefont
  {A.}~\bibnamefont {Vuorinen}},\ }\href
  {https://doi.org/10.1038/s41467-023-44051-y} {\bibfield  {journal} {\bibinfo
  {journal} {Nature Commun.}\ }\textbf {\bibinfo {volume} {14}},\ \bibinfo
  {pages} {8451} (\bibinfo {year} {2023})},\ \Eprint
  {https://arxiv.org/abs/2303.11356} {arXiv:2303.11356 [astro-ph.HE]}
  \BibitemShut {NoStop}%
\bibitem [{\citenamefont {Gorda}\ \emph {et~al.}(2026)\citenamefont {Gorda},
  \citenamefont {Komoltsev}, \citenamefont {Kurkela},\ and\ \citenamefont
  {Sunde}}]{Gorda:2025aiu}%
  \BibitemOpen
  \bibfield  {author} {\bibinfo {author} {\bibfnamefont {T.}~\bibnamefont
  {Gorda}}, \bibinfo {author} {\bibfnamefont {O.}~\bibnamefont {Komoltsev}},
  \bibinfo {author} {\bibfnamefont {A.}~\bibnamefont {Kurkela}},\ and\ \bibinfo
  {author} {\bibfnamefont {E.}~\bibnamefont {Sunde}},\ }\href
  {https://doi.org/10.3847/1538-4357/ae552a} {\bibfield  {journal} {\bibinfo
  {journal} {Astrophys. J.}\ }\textbf {\bibinfo {volume} {1002}},\ \bibinfo
  {pages} {40} (\bibinfo {year} {2026})},\ \Eprint
  {https://arxiv.org/abs/2512.18044} {arXiv:2512.18044 [astro-ph.HE]}
  \BibitemShut {NoStop}%
\bibitem [{\citenamefont {Hoogkamer}\ \emph {et~al.}(2026)\citenamefont
  {Hoogkamer} \emph {et~al.}}]{Hoogkamer:2025eaq}%
  \BibitemOpen
  \bibfield  {author} {\bibinfo {author} {\bibfnamefont {M.}~\bibnamefont
  {Hoogkamer}} \emph {et~al.},\ }\href {https://doi.org/10.1103/z2hd-w9sy}
  {\bibfield  {journal} {\bibinfo  {journal} {Phys. Rev. D}\ }\textbf {\bibinfo
  {volume} {113}},\ \bibinfo {pages} {063049} (\bibinfo {year} {2026})},\
  \Eprint {https://arxiv.org/abs/2510.27619} {arXiv:2510.27619 [astro-ph.HE]}
  \BibitemShut {NoStop}%
\bibitem [{\citenamefont {Baym}\ \emph {et~al.}(1971)\citenamefont {Baym},
  \citenamefont {Pethick},\ and\ \citenamefont {Sutherland}}]{Baym:1971pw}%
  \BibitemOpen
  \bibfield  {author} {\bibinfo {author} {\bibfnamefont {G.}~\bibnamefont
  {Baym}}, \bibinfo {author} {\bibfnamefont {C.}~\bibnamefont {Pethick}},\ and\
  \bibinfo {author} {\bibfnamefont {P.}~\bibnamefont {Sutherland}},\ }\href
  {https://doi.org/10.1086/151216} {\bibfield  {journal} {\bibinfo  {journal}
  {Astrophys. J.}\ }\textbf {\bibinfo {volume} {170}},\ \bibinfo {pages} {299}
  (\bibinfo {year} {1971})}\BibitemShut {NoStop}%
\bibitem [{\citenamefont {Hebeler}\ \emph {et~al.}(2013)\citenamefont
  {Hebeler}, \citenamefont {Lattimer}, \citenamefont {Pethick},\ and\
  \citenamefont {Schwenk}}]{Hebeler:2013nza}%
  \BibitemOpen
  \bibfield  {author} {\bibinfo {author} {\bibfnamefont {K.}~\bibnamefont
  {Hebeler}}, \bibinfo {author} {\bibfnamefont {J.~M.}\ \bibnamefont
  {Lattimer}}, \bibinfo {author} {\bibfnamefont {C.~J.}\ \bibnamefont
  {Pethick}},\ and\ \bibinfo {author} {\bibfnamefont {A.}~\bibnamefont
  {Schwenk}},\ }\href {https://doi.org/10.1088/0004-637X/773/1/11} {\bibfield
  {journal} {\bibinfo  {journal} {Astrophys. J.}\ }\textbf {\bibinfo {volume}
  {773}},\ \bibinfo {pages} {11} (\bibinfo {year} {2013})},\ \Eprint
  {https://arxiv.org/abs/1303.4662} {arXiv:1303.4662 [astro-ph.SR]}
  \BibitemShut {NoStop}%
\bibitem [{\citenamefont {Fraga}\ \emph {et~al.}(2014)\citenamefont {Fraga},
  \citenamefont {Kurkela},\ and\ \citenamefont {Vuorinen}}]{Fraga:2013qra}%
  \BibitemOpen
  \bibfield  {author} {\bibinfo {author} {\bibfnamefont {E.~S.}\ \bibnamefont
  {Fraga}}, \bibinfo {author} {\bibfnamefont {A.}~\bibnamefont {Kurkela}},\
  and\ \bibinfo {author} {\bibfnamefont {A.}~\bibnamefont {Vuorinen}},\ }\href
  {https://doi.org/10.1088/2041-8205/781/2/L25} {\bibfield  {journal} {\bibinfo
   {journal} {Astrophys. J. Lett.}\ }\textbf {\bibinfo {volume} {781}},\
  \bibinfo {pages} {L25} (\bibinfo {year} {2014})},\ \Eprint
  {https://arxiv.org/abs/1311.5154} {arXiv:1311.5154 [nucl-th]} \BibitemShut
  {NoStop}%
\bibitem [{\citenamefont {Komoltsev}\ and\ \citenamefont
  {Kurkela}(2022)}]{Komoltsev:2021jzg}%
  \BibitemOpen
  \bibfield  {author} {\bibinfo {author} {\bibfnamefont {O.}~\bibnamefont
  {Komoltsev}}\ and\ \bibinfo {author} {\bibfnamefont {A.}~\bibnamefont
  {Kurkela}},\ }\href {https://doi.org/10.1103/PhysRevLett.128.202701}
  {\bibfield  {journal} {\bibinfo  {journal} {Phys. Rev. Lett.}\ }\textbf
  {\bibinfo {volume} {128}},\ \bibinfo {pages} {202701} (\bibinfo {year}
  {2022})},\ \Eprint {https://arxiv.org/abs/2111.05350} {arXiv:2111.05350
  [nucl-th]} \BibitemShut {NoStop}%
\bibitem [{\citenamefont {Altiparmak}\ \emph {et~al.}(2022)\citenamefont
  {Altiparmak}, \citenamefont {Ecker},\ and\ \citenamefont
  {Rezzolla}}]{Altiparmak:2022bke}%
  \BibitemOpen
  \bibfield  {author} {\bibinfo {author} {\bibfnamefont {S.}~\bibnamefont
  {Altiparmak}}, \bibinfo {author} {\bibfnamefont {C.}~\bibnamefont {Ecker}},\
  and\ \bibinfo {author} {\bibfnamefont {L.}~\bibnamefont {Rezzolla}},\ }\href
  {https://doi.org/10.3847/2041-8213/ac9b2a} {\bibfield  {journal} {\bibinfo
  {journal} {Astrophys. J. Lett.}\ }\textbf {\bibinfo {volume} {939}},\
  \bibinfo {pages} {L34} (\bibinfo {year} {2022})},\ \Eprint
  {https://arxiv.org/abs/2203.14974} {arXiv:2203.14974 [astro-ph.HE]}
  \BibitemShut {NoStop}%
\bibitem [{\citenamefont {Annala}\ \emph {et~al.}(2018)\citenamefont {Annala},
  \citenamefont {Gorda}, \citenamefont {Kurkela},\ and\ \citenamefont
  {Vuorinen}}]{Annala:2017llu}%
  \BibitemOpen
  \bibfield  {author} {\bibinfo {author} {\bibfnamefont {E.}~\bibnamefont
  {Annala}}, \bibinfo {author} {\bibfnamefont {T.}~\bibnamefont {Gorda}},
  \bibinfo {author} {\bibfnamefont {A.}~\bibnamefont {Kurkela}},\ and\ \bibinfo
  {author} {\bibfnamefont {A.}~\bibnamefont {Vuorinen}},\ }\href
  {https://doi.org/10.1103/PhysRevLett.120.172703} {\bibfield  {journal}
  {\bibinfo  {journal} {Phys. Rev. Lett.}\ }\textbf {\bibinfo {volume} {120}},\
  \bibinfo {pages} {172703} (\bibinfo {year} {2018})},\ \Eprint
  {https://arxiv.org/abs/1711.02644} {arXiv:1711.02644 [astro-ph.HE]}
  \BibitemShut {NoStop}%
\bibitem [{\citenamefont {Jr.}(1951)}]{Massey01031951}%
  \BibitemOpen
  \bibfield  {author} {\bibinfo {author} {\bibfnamefont {F.~J.~M.}\
  \bibnamefont {Jr.}},\ }\href {https://doi.org/10.1080/01621459.1951.10500769}
  {\bibfield  {journal} {\bibinfo  {journal} {Journal of the American
  Statistical Association}\ }\textbf {\bibinfo {volume} {46}},\ \bibinfo
  {pages} {68} (\bibinfo {year} {1951})}\BibitemShut {NoStop}%
\end{thebibliography}
%
\end{document}